# Aether Theory Clock Retardation vs. Special Relativity Time Dilation

## Joseph Levy

4 Square Anatole France, 91250, St Germain-lès-Corbeil, France
E-mail: levy.joseph@orange.fr

Assuming a model of aether non-entrained by the motion of celestial bodies, one can provide a rational explanation of the experimental processes affecting the measurement of time when clocks are in motion. Contrary to special relativity, aether theory does not assume that the time itself is affected by motion; the reading displayed by the moving clocks results from two facts: 1/ Due to their movement through the aether, they tick at a slower rate than in the aether frame. 2/ The usual synchronization procedures generate a synchronism discrepancy effect. These facts give rise to an alteration of the measurement of time which, as we shall show, exactly explains the experimental results. In particular, they enable to solve an apparent paradox that special relativity cannot explain (see chapter 4). When the measurement distortions are corrected, the time proves to be the same in all co-ordinate systems moving away from one another with rectilinear uniform motion. These considerations strongly support the existence of a privileged aether frame. The consequences concern special relativity (SR) as well as general relativity (GR) which is an extension of SR. We should note that Einstein himself became conscious of the necessity of the aether from 1916, in contrast with conventional relativity. Yet the model of aether presented here differs from Einsein's in that it assumes the existence of an aether drift, in agreement with the discoveries of G.F. Smoot and his co-workers listed in Smoot's Nobel Lecture, December 8th 2006. Although it makes reference to previous studies, this text remains self-sufficient.

---

Version supplemented by additional information and another chapter

## 1. INTRODUCTION

In the present text, the points of view of special relativity and aether theory regarding *the measurement* of time in moving co-ordinate systems are successively presented and compared. The measurement concerns the two way transit time of light along a rod perpendicular to the direction of motion. We show that the approach of aether theory we have developed in Ref [1-4], can give a rational explanation of the experimental processes affecting this measurement, but, contrary to special relativity, these processes do not result from time dilation, but rather from the slowing down of clocks moving through the aether and from the synchronism discrepancy effect caused by the standard synchronization procedures. After correction of these measurement distortions the true value of time in moving co-ordinate systems is rediscovered. This study gives an illustration, in a specific example, of the differences existing between special relativity and aether theory.

(We should bear in mind for the reader not informed of our approach, that the concept of aether assumed in this text conforms to the Lorentz views: it is associated with a privileged aether frame and is not entrained by the motion of bodies. It is this approach that we shall refer to as "aether theory" all through the text).

This study does not question the experimental results brought about by relativity theory since, as we shall see, at least in the cases studied here, it predicts the same clock readings as SR *provided that we use the standard measurement procedures*. It nevertheless gives another interpretation of the experimental data (demonstrating that the procedures used entail measurement distortions and that the results obtained conceal hidden variables). This different interpretation and the disclosing of hidden variables should have important consequences for the future development of physics insofar as it concerns not only SR, but also GR. An important argument supporting our approach is that it solves an apparent paradox related to reciprocity that SR cannot explain (see chapter 4).

Let us bear in mind that, contrary to what is often believed, Einstein did not definitively reject the concept of aether. He assumed, no later than 1916, that the consistency of general relativity needed recognition of the aether, an opinion which he recorded in an address he delivered on May 5th 1920 in the University of Leyden [5]. But as Einstein declared at the end of this address, "the idea of motion may not be applied to this model of aether...", and, therefore, it cannot explain the discoveries of G.F. Smoot who, in a report done at the university of California, declared: "The motion of the Earth with respect to the distant matter ("aether drift") was measured, and the homogeneity and isotropy of the universe ("the cosmological principle") was probed. This recognition of an aether drift was confirmed in his Nobel lecture, December 8th 2006 [6, 7].

On the contrary, our model assumes an aether drift in agreement with the experimental studies performed by Smoot, Gorenstein and their co-workers.

## 2. TIME DILATION ACCORDING TO RELATIVITY THEORY

Let us consider two inertial coordinates systems $S_0(x, y, z)$ and $S_1(x', y', z')$ receding from one another along the *x*-axis of the co-ordinate system $S_0$, and suppose that a light ray starts from a point M fixed to the coordinate system $S_1$, and travels along a rod L=MB, perpendicular to the *x'*-axis (see fig 1). After reflection in a mirror placed in B, the signal returns to point M.

In the coordinate system $S_1$, the two-way transit time of light along the rod is $2t_1 = 2L/C$. But, viewed from $S_0$, the light ray starts from a point A in this co-ordinate system, and after reflection in B returns to point A'. The total duration of the cycle in $S_0$ will be labelled $2t_0$. According to relativity, the speed of light is *C* in all inertial frames and in all directions of space. Let $v_{01}$ refers to the real relative speed separating $S_0$ and $S_1$ (measured with non contracted standards). When the light ray has covered the distance AB, $S_1$ has moved away from $S_0$ a distance AM $= v_{01}t_0$

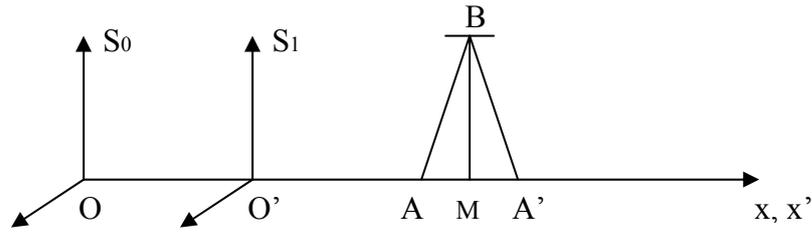

FIG 1. In the coordinate system $S_1$, the light ray travels from point M to the mirror B and, after reflection, returns to M. In $S_0$ the signal starts from point A, and after reflection in B returns to A'.

According to an observer attached to $S_0$ the transit time of light $t_0$ along AB is given by: $C^2 t_0^2 - v_{01}^2 t_0^2 = L^2$, therefore:

$$t_0 = \frac{L}{C\sqrt{1 - v_{01}^2 / C^2}}.$$

Replacing *L/C* by its value $t_1$ this expression reduces to:

$$t_0 = \frac{t_1}{\sqrt{1 - v_{01}^2 / C^2}}.$$

This classical formula is interpreted as time dilation by special relativity, (an expression obtained because the position of the clock in $S_1$ remains fixed relative to this co-ordinate system)

### 3. CLOCK RETARDATION ACCORDING TO AETHER THEORY

The theory on which this study is based, assumes the existence of a preferred frame in which the aether is at rest. The one-way speed of light is $C$ in the aether frame, and different from $C$ in all other co-ordinate systems moving with respect to the aether frame. Yet, as we saw in Ref [1-4] and [8], due to measurement distortions (that will be evoked in the text which follows), it appears to be of magnitude $C$ in all 'inertial' frames and in all directions of space.

Contrary to relativity theory, the motion of bodies does not affect the time, but the motion through the aether causes a slowing down of the moving clocks. The real two-way transit time of light, along a rod attached to a certain 'inertial' frame, is the same for the observers of all frames, but, due to clock retardation, the reading displayed by clocks moving relative to the rod will depend on their speed with respect to the rod [2, 8, 9].

Although the variables used in this study should be difficult to determine experimentally, our approach, as we shall see, allows an exact theoretical comparison of the concepts of time assumed by the two theories.

In this section we shall study successively two different cases: in section *3.1.* the clock reading in a moving 'inertial' co-ordinate system is compared to the time in the aether frame; this case introduces to the section *3.2.* which puts forward exhaustively the differences between aether theory and relativity.

The paradox inherent in conventional relativity when we assume a complete symmetry between frames will be examined in section 4.

*3.1. Comparison of the clock readings displayed in frames $S_0$ and $S_1$*

In this section we shall compare the clock readings in two 'inertial' co-ordinate systems as we did in section **2**, but from the point of view of aether theory. The only difference is that the co-ordinate system $S_0$ is assumed to be at rest in the aether frame where the clock reading is not altered by motion (and which can be regarded as the basic time or, by definition, the real time) (Fig 1). Since the line AB is the path of the light signal in the aether frame, the speed of light is $C$ along this line.

Referring to the transit time of light along AB, that would be displayed by a clock attached to frame $S_0$, as $t_0$, we have:

$$C^2 t_0^2 - v_{01}^2 t_0^2 = L^2,$$

and therefore:

$$t_0 = \frac{L}{C\sqrt{1 - v_{01}^2/C^2}}.$$

According to the aether theory under consideration, clock retardation is defined with respect to the aether frame; the ratio between the time in the aether frame and the reading displayed by clocks moving at absolute speed $v$ *is* assumed to be equal to $(1 - v^2/C^2)^{-1/2}$. This assumption will be justified a posteriori; its experimental implications will be studied in the text that follows.

Therefore, the clocks attached to the co-ordinate system $S_1$ tick at a slower rate and display the reading $t_{1app} = L/C$.

Thus,

$$t_0 = \frac{t_{1app}}{\sqrt{1 - v_{01}^2/C^2}}, \qquad (1)$$

where the suffix 'app' means apparent.

This formula assumes the same mathematical form as the time dilation formula of special relativity; yet its meaning is quite different because, contrary to special relativity, $t_{1app}$ is not the true time in the co-ordinate system $S_1$, it is the clock reading displayed by clocks slowed down by motion.

(We bear in mind that, if we assume the existence of a preferred aether frame, then, real frames attached to bodies, even if they are not submitted to physical influences other than the aether drift, are never perfectly inertial. The term 'inertial' is an approximation which must be limited to the cases where the absolute speed of the frames under consideration is low compared to the speed of light. See Ref [2]).

*3.2. Case of two co-ordinate systems moving away from the aether frame*

We now propose to study a different case: we shall determine the clock retardation formula between two co-ordinate systems $S_1$ and $S_2$ receding with rectilinear uniform motion with respect to the co-ordinate system $S_0$ which is attached to the aether frame. The direction of motion is the *x*-axis (see fig 2). This case is that to which we usually deal with in practice.

In relativity, there is no preferred frame, therefore the co-ordinate system $S_0$ is inexistent and the time dilation formula between the systems $S_1$ and $S_2$ takes the form:

$$T_1 = \frac{t_2}{\sqrt{1 - v_{12}^2/C^2}}, \qquad (2)$$

where $v_{12}$ refers to the relative speed between the coordinate systems $S_1$ and $S_2$.

In aether theory, things are very different. Let $v_{01}$, $v_{02}$ and $v_{12}$ refer to the real relative speeds between the three co-ordinate systems (obtained in the absence of measurement distortions). The rod MB perpendicular to the *x"*-axis is firmly fixed to the co-ordinate system $S_2$. We propose to compare the *apparent* times (displayed by the clocks attached to $S_1$ and $S_2$) which are needed by the light signal to achieve a cycle (from M to B and to M again in the system $S_2$ and from A to B and to A' in the system $S_1$).

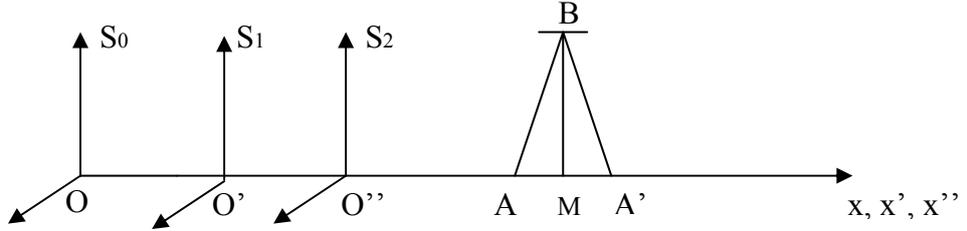

FIG 2. The co-ordinate systems $S_1$ and $S_2$ recede from $S_0$ along the common *x*-axis. The light ray travels along the rod MB which is at rest in $S_2$ (from M to B and to M again). With respect to $S_1$ it starts from point A, is reflected in B and then returns to A', (where A and A' are two points at rest in the co-ordinate system $S_1$). During a cycle of the signal, $S_2$ has moved with respect to $S_1$ a distance AA'. The real value of the one-way speed of light along AB is not equal to $C$ since the coordinate system $S_1$ is not at rest with respect to the aether frame.

*3.2.1. We shall first assume that the clocks placed at points A and A' are exactly synchronized.* Let us label as $2t_0$ the two-way transit time of the light signal that would be displayed by clocks attached to the co-ordinate system $S_0$.

Due to clock retardation the clock readings in $S_1$ and $S_2$ are related to $t_0$ as follows:

$$t_{1app} = t_0\sqrt{1 - v_{01}^2/C^2} \;, \qquad (3)$$

and

$$t_{2app} = t_0\sqrt{1 - v_{02}^2/C^2} \;. \qquad (4)$$

(Note nevertheless that the true time, needed for half a cycle, measured with clocks not slowed down by motion, is $t_0$ for all observers).

From (3) and (4) we infer:

$$t_{1app} = t_{2app} \frac{\sqrt{1 - v_{01}^2/C^2}}{\sqrt{1 - v_{02}^2/C^2}} \;. \qquad (5)$$

Assuming that $v_{02}/C \ll 1,$ this expression reduces to first order, to:

$$\frac{t_{2app}}{1-\frac{1}{2}\frac{v_{12}}{C^2}(v_{12}+2v_{01})} \quad . \tag{6}$$

Noting that (as in section *3.1.*) $t_2 = t_{2app} = L/C$, this expression is different from the relativistic formula (2) which reduces to:

$$\frac{t_2}{1-\frac{1}{2}\frac{v_{12}^2}{C^2}} \quad . \tag{7}$$

Therefore if clocks were exactly synchronized, there would be an obvious difference between the two theories.

*3.2.2. Practical consequences of the clock synchronization procedures used.* We should note that, in practice, in order to determine the duration of a cycle in $S_1$ we must subtract the reading displayed by clock A when the signal starts from this clock, from the reading displayed (after reflection in B) by clock A' when the signal reaches this clock, and therefore we must synchronize the clocks A and A' beforehand.

According to aether theory if the synchronization of clocks was perfect we would have obtained formula (5). Yet, synchronizing the clocks perfectly is a difficult problem, and, with the standard synchronization procedures, (Einstein-Poincaré method (E. P) or slow clock transport), we make an unavoidable systematic error in measuring the time, (synchronism discrepancy effect) [2, 9, 10]

The *apparent* duration of a cycle measured in $S_1$ is therefore equal to the difference between $2t_{1app}$ and the synchronism discrepancy effect (SDE) that will be derived in the text which follows. (The SDE, which was defined by Prokhovnik for the first time, enables to resolve a number of paradoxes in physics).

Referring to the SDE that would affect the clocks if they were not slowed down by motion as $\Delta$, the SDE affecting the clocks attached to the coordinate system $S_1$ is:

$$\delta = \Delta\sqrt{1-v_{01}^2/C^2} \ .$$

The *apparent* (measured) two-way transit time of the signal (from A to B and to A') is therefore:

$$2T_{1app} = 2t_{1app} - \Delta\sqrt{1-v_{01}^2/C^2} \ .$$

It is this apparent time which is in fact measured when a SDE between the clocks A and A' exists.

(a) *Derivation of the synchronism discrepancy effect, and clock synchronization.* The Einstein-Poincaré method (E. P) consists in sending a light signal from clock A to

- 7 -

clock A' along the *x'*-axis, at an arbitrary instant where the reading of clock A is set at *t*=zero. After reflection in A', the signal comes back to A. The clocks are considered synchronous if upon reception of the signal by clock A', this clock displays a reading equal to half the reading displayed by clock A upon return of the signal.

(The alternative synchronization method, referred to as the slow clock transport procedure, has been shown to be equivalent to the former by different authors [2, 10]).

Although the measurement should be difficult to perform with our today technology, it is possible to carry out a theoretical evaluation, as we shall see, of how the SDE modifies the reading of the time (in comparison with the clock readings displayed by clocks exactly synchronized given by formula (5)). The result will then be compared to the time dilation of special relativity, a comparison that will enable to check the theory.

Let us label as $\ell_0$ the length that would be assumed by the segment AA' if it was at rest in the aether frame. Due to its motion with respect to $S_0$ it is reduced to

$$\ell = \ell_0 \sqrt{1 - v_{01}^2 / C^2} \; , \tag{8}$$

which, according to aether theory, is the real length in the co-ordinate system $S_1$.

The real time needed by the light signal to travel from A to A' along the x'-axis is therefore:

$$t_{raa'} = \frac{\ell_0 \sqrt{1 - v_{01}^2 / C^2}}{C - v_{01}} \; ,$$

where $C - v_{01}$ is assumed to be the real speed of light in $S_1$ along the *x'*-axis. Here the suffix r (for real) means that the determination of the speed is made without measurement distortions. This formula was the expression used by Lorentz to explain the Michelson experiment. (According to aether theory, real speeds measured along a straight line are simply additive. Only *apparent* speeds (whose measurement is altered by the systematic measurement distortions) obey the relativistic law of composition of velocities, as we shall see in formula (14). (See also Refs [1, 2]).

In the reverse direction we have:

$$t_{ra'a} = \frac{\ell_0 \sqrt{1 - v_{01}^2 / C^2}}{C + v_{01}} \; .$$

Half the two way transit time of the light signal along the *x'*-axis (from A to A' and to A again) measured with clocks not slowed down by motion is therefore:

$$1/2(t_{raa'} + t_{ra'a}) = \frac{\ell_0}{C\sqrt{1 - v_{01}^2 / C^2}} \; .$$

In the absence of clock retardation, the synchronism discrepancy $\Delta$ between the clocks A and A' would be equal to the difference between the exact transit time $t_{raa'}$ of the signal from A to A' and the *apparent* (measured) time $1/2(t_{raa'} + t_{ra'a})$:

$$\Delta = \frac{\ell_0\sqrt{1-v_{01}^2/C^2}}{C-v_{01}} - \frac{\ell_0}{C\sqrt{1-v_{01}^2/C^2}} = \frac{v_{01}\ell_0}{C^2\sqrt{1-v_{01}^2C^2}}.$$

Due to clock retardation in the co-ordinate system $S_1$ the SDE is reduced to

$$\delta = \frac{v_{01}\ell_0}{C^2}.$$

(b) *Apparent transit time of light along the rod in the co-ordinate system $S_1$*. In the absence of SDE, the *apparent* transit time of light from A to B and to A' again measured with clocks slowed down by motion would be:

$$2t_{1app} = 2t_0\sqrt{1-v_{01}^2/C^2}.$$

If one takes account of the SDE, the clock reading becomes

$$2T_{1app} = 2t_{1app} - \frac{v_{01}\ell_0}{C^2}. \tag{9}$$

<u>Important remark</u>
Writing this expression in the form

$$2T_{1app} = 2t_0\sqrt{1-v_{01}^2/C^2} - \frac{v_{01}\ell_0}{C^2}, \tag{10}$$

and taking account of the fact that the measurements in $S_1$ are made with a meter stick which is also contracted, the length AA' is erroneously found equal to $\ell_0$. We shall therefore refer to $\ell_0$ as $X_{1app}$.
From (10) we obtain:

$$2t_0 = \frac{2T_{1app} + v_{01}X_{1app}/C^2}{\sqrt{1-v_{01}^2/C^2}}. \tag{11}$$

This expression assumes the same mathematical form as the conventional transformation relative to time, yet its meaning is quite different since it demonstrates that the variables $2T_{1app}$ and $X_{1app}$ which are obtained experimentally are different from the true values. The experiment is altered by measurement distortions [1].
(Such a mathematical form is obtained because, contrary to the case studied in section *3.1.*, the measurement of time in $S_1$ is made in two different points of the *x'-axis*).

-Taking account of expression (8) and of the fact that $\ell = (v_{02} - v_{01}) \times 2t_0$ where $v_{02}$ and $v_{01}$ are the real speeds of $S_2$ and $S_1$ with respect to $S_0$ (obtained in the absence of measurement distortions) we can express $\delta = \dfrac{v_{01}\ell_0}{C^2}$ in the form

$$\delta = \frac{2v_{01}t_{1app}(v_{02} - v_{01})}{C^2 - v_{01}^2}.$$

The *apparent* (measured) transit time of the signal in $S_1$ (from A to B and to A' again) is therefore:

$$2T_{1app} = 2t_{1app} - \delta = 2t_{1app}\frac{C^2 - v_{01}v_{02}}{C^2 - v_{01}^2}.$$

Now our objective is to compare the *apparent* (measured) transit times of the signal in frames $S_1$ and $S_2$. From formula (5) we have:

$$2T_{1app} = 2t_{2app}\frac{\sqrt{1 - v_{01}^2/C^2}}{\sqrt{1 - v_{02}^2/C^2}}\frac{C^2 - v_{01}v_{02}}{C^2 - v_{01}^2}. \tag{12}$$

From formula (12) we obtain successively:

$$2T_{1app} = 2t_{2app}\frac{C^2 - v_{01}v_{02}}{C^2\sqrt{(1 - v_{02}^2/C^2)(1 - v_{01}^2/C^2)}}$$

$$= 2t_{2app}\frac{C^2 - v_{01}v_{02}}{\sqrt{C^4 - v_{02}^2 C^2 - v_{01}^2 C^2 + v_{01}^2 v_{02}^2}}$$

$$= 2t_{2app}\frac{C^2 - v_{01}v_{02}}{\sqrt{C^4 + v_{01}^2 v_{02}^2 - 2C^2 v_{01}v_{02} - v_{02}^2 C^2 - v_{01}^2 C^2 + 2C^2 v_{01}v_{02}}}$$

$$= 2t_{2app}\frac{C^2 - v_{01}v_{02}}{\sqrt{(C^2 - v_{01}v_{02})^2 - C^2(v_{02} - v_{01})^2}} = \frac{2t_{2app}}{\sqrt{1 - \dfrac{C^2(v_{02} - v_{01})^2}{(C^2 - v_{01}v_{02})^2}}}$$

$$2T_{1app} = \frac{2t_{2app}}{\sqrt{1 - \dfrac{(v_{02} - v_{01})^2}{C^2(1 - \dfrac{v_{01}v_{02}}{C^2})^2}}}. \tag{13}$$

<u>This result is different from formula (5) which was the exact clock retardation formula according to aether theory. Yet, it is the experimental formula obtained in practice due to the measurement distortions</u>.

We recognize in the denominator the experimental composition of velocities law. Since it has been derived from the Galilean law which has been submitted to measurement distortions its apparent character is highlighted. We can thus write:

$$v_{12app} = \frac{v_{02} - v_{01}}{1 - v_{01}v_{02}/C^2} . \qquad (14)$$

With this notation formula (13) becomes:

$$T_{1app} = \frac{t_{2app}}{\sqrt{1 - v_{12app}^2/C^2}} . \qquad (15)$$

This formula has the same mathematical form as the time dilation formula (2) of relativity theory, but obviously its meaning is quite different. In particular we have not assumed the invariance of the one-way speed of light in all inertial frames.
This surprising result cannot be the effect of chance.
(Note that this formula assumes a mathematical form different from formula (11). This difference results from the fact that in the co-ordinate system $S_2$ the time is measured at the same point at the beginning and at the end of a cycle).

We should note that, when the co-ordinate system $S_1$ is at rest in the preferred aether frame, $v_{01} = 0$ and $v_{12app}$ reduces to $v_{02}$. This demonstrates that contrary to what is often claimed, the aether frame can be theoretically distinguished from the other frames. This result, which is in accordance with the experimental facts, strongly supports the existence of a privileged aether frame in a state of absolute rest.

## 4. THE QUESTION OF RECIPROCITY (RESOLUTION OF A PARADOX).

The resolution of the paradoxes inherent in reciprocity which affect special relativity have been first suggested by Builder and Prokhovnik [9].
We will consider here the paradox affecting the measurement of time from the device described in the previous chapters.
Let us return to the figure 1. In the case studied there, the rod MB was at rest with respect to the co-ordinate system $S_1$ and was moving relative to the co-ordinate system $S_0$. We shall now consider the opposite case: i.e. the rod is at rest with respect to $S_0$ and the co-ordinate system $S_1$ moves relative to $S_0$ in the left direction, as the figure 3 shows.

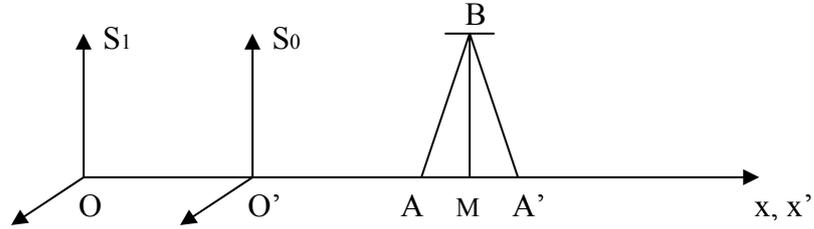

FIG 3. The rod L=MB is at rest with respect to the co-ordinate system $S_0$, and the co-ordinate system $S_1$ is moving relative to $S_0$ in the left direction.

According to conventional relativity, contrary to aether theory, nothing differentiates the co-ordinate systems $S_0$ and $S_1$, because there is no preferred inertial frame; in other words, motion is only relative, and one can consider that $S_0$ moves relative to $S_1$, in the same way as $S_1$ is moving relative to $S_0$.
Therefore SR predicts a complete symmetry between the frames: for example, a clock in $S_1$ slows down with respect to a clock standing in $S_0$, but conversely a clock in $S_0$ is subjected to slow down with respect to a clock in $S_1$. Of course this result appears paradoxical. It defies logic and cannot be rationally explained if this total equivalence between frames is assumed.
Yet, as we shall see, the paradox can be solved if we assume the existence of a preferred aether frame in which case the measurements are affected by systematic distortions, and the complete symmetry proves only apparent.
The *apparent* identity of the two opposite situations results from these measurement distortions as the following demonstration will show.

In agreement with aether theory, let us assume that $S_0$ is a co-ordinate system at rest in the preferred frame, and $S_1$ a co-ordinate system moving at uniform speed in the left direction, see Fig 3.
Since it is $S_1$ which moves, the clocks in $S_1$ tick slower than in $S_0$ and we should have:

$$t_1 = t_0\sqrt{1 - v_{01}^2/C^2} = \frac{L}{C}\sqrt{1 - v_{01}^2/C^2} . \qquad (16)$$

Where $v_{01}$ is the real speed between $S_0$ and $S_1$, (measured with non-contracted meter sticks and clocks non slowed down by motion)

Yet, this result supposes that the measurement has been made with clocks exactly synchronized, and as we saw, the exact synchronization is an objective difficult to achieve and which is not practiced today.

Let us describe theoretically the method which therefore should be used by an observer at rest in $S_1$ in order to measure the two way transit time of light along the rod MB, by means of the E. P procedure.

Viewed from the co-ordinate system $S_0$, the light ray travels along the rod MB (from M to B and to M again), but viewed from $S_1$, it starts from A, is reflected in B and comes back to A', where A and A' are two points at rest with respect to the co-ordinate system $S_1$.

In the absence of length contraction let us suppose that AA' measures $\ell_0$. Taking account of the reduction of size, we have:

$$\text{AA'} = \ell = \ell_0 \sqrt{1 - v_{01}^2 / C^2} = 2 v_{01} t_0 .$$

According to the E. P procedure, the synchronization requires two clocks placed in A and A'. The clocks are considered synchronous if, when a light ray starting from A at the instant zero and travelling along the x, x'-axis strikes A', this clock displays the reading:

$$1/2 \; \ell \; (\frac{1}{C + v_{01}} + \frac{1}{C - v_{01}}) \sqrt{1 - v_{01}^2 / C^2} = \frac{\ell}{C \sqrt{1 - v_{01}^2 / C^2}} = \frac{\ell_0}{C} .$$

(In this expression we have taken account of the slowing down of the clocks standing in the co-ordinate system $S_1$, and we have made use of the Galilean composition of velocities law which applies in aether theory to real speeds).

Note

Let us remark that in $S_1$ the standard used to measure the lengths is contracted in the same ratio as the segment AA'. Therefore the length AA' is erroneously found equal to $\ell_0$ and therefore the light speed is erroneously found equal to C in conformity with the experiment.

Yet the real transit time of the light ray from A to A' along the x, x'-axis is:

$$\frac{\ell_0 \sqrt{1 - v_{01}^2 / C^2}}{C + v_{01}} .$$

But due to the slowing down of clocks in the co-ordinate system $S_1$, the reading in the absence of synchronism discrepancy would be:

$$\frac{\ell \sqrt{1 - v_{01}^2 / C^2}}{C + v_{01}} .$$

The clock A' will therefore be ahead of the clock A by an amount equal to:

$$\frac{\ell}{C\sqrt{1-v_{01}^2/C^2}} - \frac{\ell\sqrt{1-v_{01}^2/C^2}}{C+v_{01}} = \frac{v_{01}\ell}{C^2\sqrt{1-v_{01}^2/C^2}} = \frac{2v_{01}^2 t_0}{C^2\sqrt{1-v_{01}^2/C^2}}.$$

The measured time of light transit from A to B and to A' again being made with the clocks A and A', the result of the measurement will give (instead of $2t_1 = 2t_0\sqrt{1-v_{01}^2/C^2}$

$$2t_{1app} = 2t_0\sqrt{1-v_{01}^2/C^2} + \frac{2v_{01}^2 t_0}{C^2\sqrt{1-v_{01}^2/C^2}},$$

Which yields
$$t_{1app} = \frac{t_0}{\sqrt{1-v_{01}^2/C^2}} \qquad (17)$$

This clock reading is the *apparent* time resulting from the synchronism discrepancy effect. This result enables to explain rationally *the paradoxical effect which is anticipated by special relativity without being explained*. Yet special relativity regards this result as the true time, while it is in fact an *apparent* time resulting from the measurement distortions.

**5. CONCLUSION**
The comparison of formulas (6) and (7) demonstrates that relativity and aether theory are fundamentally different. Nevertheless, paradoxically, due to the systematic measurement distortions mentioned above, aether theory leads to a clock reading given by formula (15), which presents a mathematical form identical to formula (2); yet for relativity, the formula is regarded as exact, while for aether theory it results from the measurement distortions.
Aether theory provides also an explanation of why formulas (1) and (17) can be both rationally justified, although at first sight they appear incompatible. Aether theory explains that due to the synchronism discrepancy effect formula (17) is observed instead of formula (16), an explanation which solves the paradox.
 Special relativity obtains the same result but cannot give a rational explanation of it.
In conclusion, the choice of one theory rather than the other is not simply a question of philosophical preference.

The ideas expressed in this article are identical to those which were developed in the previous version published in arXiv (*Physics/*0611077). We have only given further explanations and added the chapter 4.

**ACKNOWLEDGEMENTS**
I would like to thank Pr. Gianfranco Spavieri and Dr. Dan Wagner for interesting and helpful comments.